# Sensitivity in Nanomechanical Pedestal MEMS Cantilever

Abhay K. Rajak[1], Ritambhara Dash[1], Ashwini Kumari[1], A.S. Bhattacharyya[1, 2 a]

[1] Department of Metallurgical and Materials Engineering, Central University of Jharkhand, Ranchi, India, 835205
[2] Centre of Excellence in Green and Efficient Energy Technology (CoE GEET), Central University of Jharkhand, Ranchi 835205, India

[a] Corresponding author Email: arnab.bhattacharya@cuj.ac.in

## Abstract

Nanomechanical resonator-based sensing devices are used in medical diagnostics based on their high-frequency dynamic behavior. Cantilevers fall into the category of Nanomechanical resonators. It also resembles a resonator whose shape is like that of a nanowire clamped at one end. As the surface-to-volume ratio of a nanowire resonator increases due to scaling down, surface stress plays a crucial role in the mechanical behavior of a resonator. Piezoresistive MEMS cantilevers are used for vapor phase analysis of volatile compounds and gas. Studies were done to address the mass sensitivity issues and fractures associated with bioceramic and nanocomposite coatings-based cantilever resonators. The studies show how the sensing performance can be determined or tuned. Nanomechanical studies of thin films of SiCN on silicon were performed. The sharpness of the tip was found to have an influence on the tip-sample conduction mechanism useful for MEMS applications

## 1. INTRODUCTION

Due to several benefits, such as their simple geometry (such as a Cantilever beam structure), the ability to use batch fabrication, their suitability for extreme miniaturization, even in the nanoscale, and their high mass sensitivity, MEMS-based resonant sensors have been widely used as biological, physical, and chemical sensors **[1, 2]**. As the sensor's resonant frequency, $\omega$, is inversely correlated to the square root of its total mass, ($\omega_o \sim 1/\sqrt{m}$) observation of the resonant frequency shift between the system with and without the target mass yields the target entity's mass. Utilizing this feature of the MEMS-based resonator, measurements of mass, stiffness, viscosity, and other physical properties have been made.

Research into cell biology, tissue engineering, cancer, and diseases may benefit greatly from the characterization of the physical characteristics of living cells, such as their mass and stiffness **[3].** The ability to measure the physical characteristics of cells provides the chance to solve unanswered issues about the development of biological systems. For instance, the mechanisms underpinning cell cycle progression can be clarified by examining the direct relationship between cell growth rate and cell mass for individual adherent human cells. **[4, 5].** A MEM-based resonator is one of the frequently used instruments to measure the mass and stiffness of the individual cell. There are numerous techniques to measure these parameters, and one of the widely-used devices to measure these quantities is a MEM-based resonator **[2].** Thin films memristor from hBN and SiC show good binary resistive memory switching which is beneficial for memory devices. Silicon-based piezoelectric nano/microelectromechanical systems (N/MEMS) are being integrated with memristors. The nano resistive switch properties can be studied with the help of nanoindentation as reported for amorphous (a-SrTiO3) perovskites memristors **[6]**.

In this communication, Si-based cantilever MEMS cantilevers were fabricated using lithographic techniques, and the issue of mass sensitivity associated with Cantilever MEMS was addressed Finite Element Analysis of a novel design has been provided. As Silicon-based MEMS sensors have the limitation of operating at high temperatures high temperature-resistant multicomponent hard coatings were deposited. The nanomechanical characterization of these coatings using nanoindentation techniques gave insights into their mechanical strength and toughness.

## 2. EXPERIMENTAL PROCEDURES

The fabrication process of Si-based Cantilever MEMS was done in a clean room with proper aprons, head covers, face masks, and foot covers following the safety protocols. RCA cleaning of a silicon wafer took place in *Wet Etch Bay*. RCA 1 cleaning is used for the organic contaminants whereas RCA 2 is done for the metallic contaminants. The RCA 1 cleaning is equivalent to piranha ($H_2SO_4$ and $H_2O_2$) cleaning for removing organic contaminants. The samples were also rinsed in DI water and blow-dried with nitrogen gas.

Silicon Nitride ($Si_3N_4$) was deposited on the silicon substrate by Low-Pressure Chemical Vapour Deposition (LPCVD) in the Diffusion Bay. The equipment had three chambers: *Loading*: where the samples were loaded in a boat made up of quartz and heated to 500oC. Nitrogen gas was passed and the temperature was further raised to 850 $^o$C. In the *process* chamber, 120 sccm of dichlorosilane (DSC) and ammonia ($NH_3$) at 300 mTorr were supplied to deposit silicon nitride. After the deposition, the sample was transferred to the *unloading* chamber and the temperature was lowered again to 500oC by applying nitrogen gas. This Si3N4 thin film formed is called *low-stress* silicon nitride deposition. Isi-rich rich and mainly used for cantilevers where stiction needs to be avoided. An ammonia-rich high-stress silicon nitride can also be prepared to take 10 sccm of DSC and 70 sccm of ammonia. The LPCVD can operate between 200 – 500 mTorr and the temperature can go up to 950 $^o$C. There are four tubes in the LPCVD system: in Tube 1 a reaction with dichlorosilane ($SiCl_2\ H_2$) with ammonia take place to form silicon nitride as discussed above. In Tube 2, thin films of Si, Ge, SiGe, a-Si Si- Nanowires, Ge, etc can take place. In this tube some metal contamination is allowed however in Tube 3 no metal contamination is allowed with the same depositing materials as in tube 2. Low-temperature oxide deposition of average 100 nm thickness using silane and oxygen as precursors takes place in Tube 4. Oxidation and diffusion furnaces were also present which can perform dry oxidation (5nm – 150 nm) and pyrogenic oxidation (250 nm – 1 μm). Phosphorous and Boron diffusion in Si for doping purposes can also be done.

The thickness of the $Si_3N_4$ thin film was estimated to be around 231 nm in ellipsometry. The thickness measurement is done on the principle of polarization. The film was also found to be an absorbing one with a refractive index of 2. Lithography was then performed as per the following procedures: Substrate cleaning (in acetone and IPA) followed by dehydration bake. A positive photoresist was then coated using spin coating followed by a soft bake. UV exposure was done using a mask designed for cantilevers. The etching can be isotropic or anisotropic. Anisotropic etching corresponds to different etch rates at different crystallographic planes. The chemical wet etch is usually an isotropic etch whereas the as dry etch is usually an anisotropic etch. The reason for making $Si_3N_4$ films rather than SiO2 films is: $SiO_2$ unlike $Si_3N_4$ is not a good mask for longer etch duration.

The mask used in lithography is made up of soda-lime glass, chrome, and, photoresist. While for labeling the masks more than one label should be used otherwise any rotation may get undetected. Dry etching was using $SF_6$ which is an isotropic etching as bombarding as well as the chemical reaction takes place. The resist was then removed by $O_2$ plasma – a process called PR ashing. Dry etching consists of Plasma generation (isotropic), sputtering, reactive ion etch (RIE) and Deep Reactive Ion etch (DRIE). RIE and DRIE are anisotropic etching processes. DRIE is done with C4F8(Scallop) and can etch up to 400 – 500 μm. E-beam lithography was shown which uses a PMMA photoresist. The apparatus consists of a laser interferometer is used. The aperture size varies between 7.5 μm to 120 μm. It uses an accelerating voltage of 30 kV producing high-energy electrons which undergo less scattering. Au is used as identification marks. The microscopic inspection was done showing the *release* of the Cantilevers. An image of the fabricated cantilever is shown in **Fig 1**.

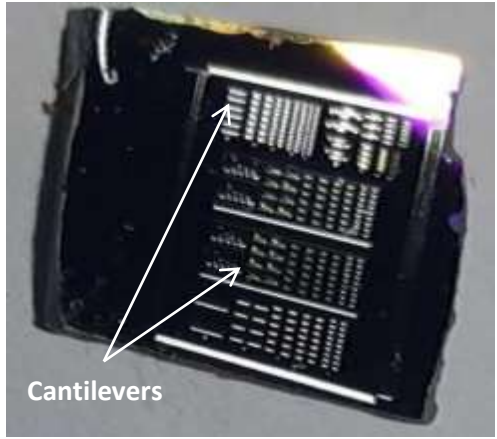

**Fig 1**. Fabricated Si-based Cantilever structure (at CENSE, IISc Bangalore)

Silicon is not able to withstand high temperatures and is being coated with hard nanocomposite coatings like SiCN which itself has potential applications in MEMS. The parameters of SiCN useful for piezoresistive sensing applications are given in **Table 1**. The deposition was carried out using rf magnetron sputtering using a sintered SiC target in an evacuated chamber with a combination of Ar/$N_2$ gas, the details of which have been published previously along with extensive structural and mechanical characterization **[7-10]**. A sintered SiC pellet was used as the target and Nitrogen gas along with Argon was introduced in the evacuated chamber for SiCN deposition on Silicon, details which have been published previously **[11]**. Nanoindentation was done on the SiCN films by MTS (USA) nanoindenter. The mechanism of nanoindentation is given in detail in ref **[12, 13].** The nanoindentation tests were performed by MTS Nanoindenter, USA having a 3-sided pyramidal Berkovich indenter based on continuous stiffness mode (CSM) **[14].**

Table 1. Parameters of SiCN useful for piezoresistive applications

| Parameter | Value |
|---|---|
| Bandgap | 2.3–3.0 eV |
| Break down voltage | 29 V at RT with leakage current density $1.2 \times 10^{-4}$ A/cm$^2$ |
|  | 5 V at 200oC $1.47 \times 10^{-4}$ A/cm$^2$ |
| Modulus | 240 GPa |
| Chemical inertness | excellent |
| MEMS compatibility | excellent |

## 3. RESULTS & DISCUSSIONS

### 3.1 MEMS cantilever structure analysis

Cantilever-based MEMS resonator sensors suffer the problem of un attainment of homogenous mass sensitivity. The intensity of vibration being highest at the free end of the cantilever makes the mass sensitivity highest and reduces as one moves towards the fixed end. The resonant frequency of operation is inversely proportional to the mass., and impulse plays a very significant role in these devices. The schema f spatially non-uniform mass sensitivity of greater than 100% from the free end of the cantilever to the middle of the cantilever can be found in **Fig 2(a).** A novel design with pedestal geometry has been proposed which can solve this problem **[4, 15]**. The resonator consists of a square pedestal suspended by four beam springs as shown in **Fig 2(b)**. The pedestal oscillates in the vertical direction while vibrating, and the end of four beam springs is fixed to the substrate. **Fig 2(b)** also shows the pedestal design with spatial non-uniformity of mass sensitivity to be less than 4% from the center to the edge of the platform. The difference occurs due to radial variation as compared to linear variation as in the previous case. The color distribution indicates the sensitivity with red being the most sensitive and blue being the least sensitive. **Fig 2(c)** shows the arrangement of the pedestal cantilever as an array for use as a sensor **[4]**.

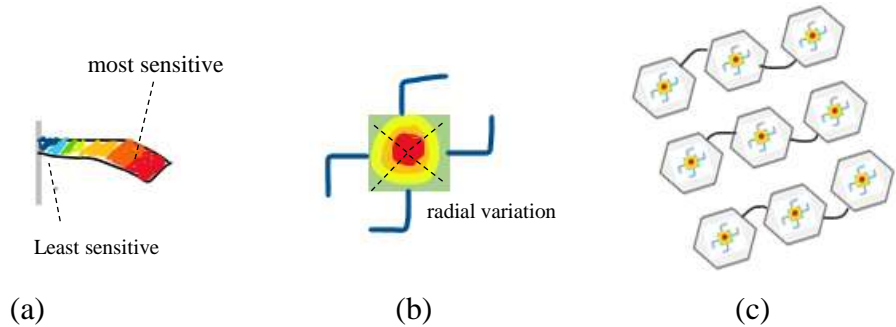

(a)  (b)  (c)

**Fig 2.** Schema of (a) spatially non-uniform mass sensitivity of greater than 100% from the free end of the cantilever to the middle of the cantilever.(b)The Pedestal design with spatial non-uniformity of mass sensitivity to be less than 4% from the center to the edge of the platform. The color distribution indicates the sensitivity with red being the most sensitive and blue being the least sensitive (c) Pedestal sensor array**[4]**

Finite element analysis is performed (ANSYS 2022 R12, ANSYS Inc.) on this novel design using the geometry and mesh distribution, and the directional deformation, total deformation, and minimum principal strain values were determined (**Fig 3**) The mesh distribution is given in Fig 3(a). The directional deformation in **Fig 3(b)** and the total deformation in Fig 3(c) are responses of the pedestal structure to the stresses imposed on it due to the capture of biological molecules showing spatial non - uniformity existing in a linear fashion for the first case diagonally for the second case. The pedestal structure is however arranged in an array for sensing as shown above which nullifies the effect as one pedestal cantilever compensating for the other.  The minimum principle strain was found to be homogenous as well as of higher proportions (red region throughout the surface) indicating the superiority of these structures compared to the conventional cantilevers with one fixed end.

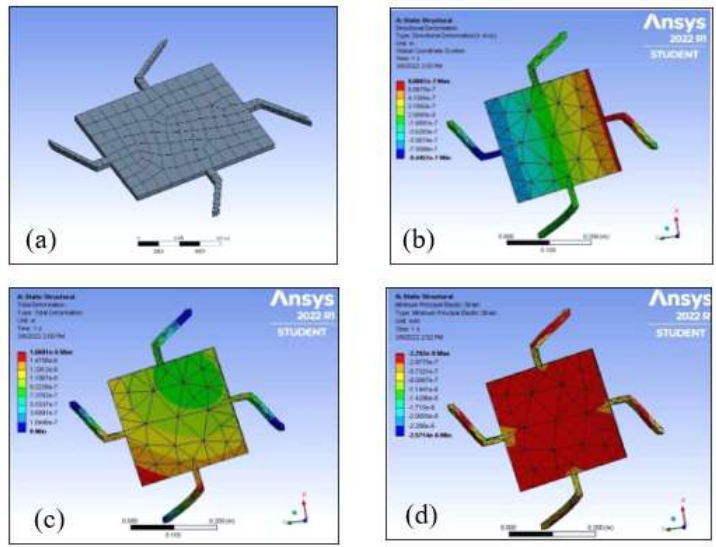

**Fig 3** a) Geometry of pedestal novel design with mesh distribution b) Directional deformation c) Total deformation d) Minimum principal strain

### 3.2 Nanomechanical studies

A hardness of 11 GPa and a Modulus of 140 GPa were indicating the good mechanical properties of the deposited films (**Fig 4 b**). Hysteresis during unloading was observed during unloading due to pressure-induced phase transformation occurring in Si-C-N/Si films which can be related to switching properties in memristors (**Fig 4 a**). As the films were about 50 nm thick, the influence of the Si substrate was prominent. Thicker coatings usually do not show this hysteresis effect. An increase in load caused an increase in conduction due to increased defect densities and diffusion of oxygen vacancies. The formation of conductive channels takes place by extension of defect structures. The hysteresis effect has been found in 2D materials and transparent conductive multilayer films due to imperfect elastic behavior. The evidence of thermally-induced transformations has been observed through partial indent recoveries at the nanoscale. Nanoindentation has also been used in for magnetoelectric memory devices **[16-19].**

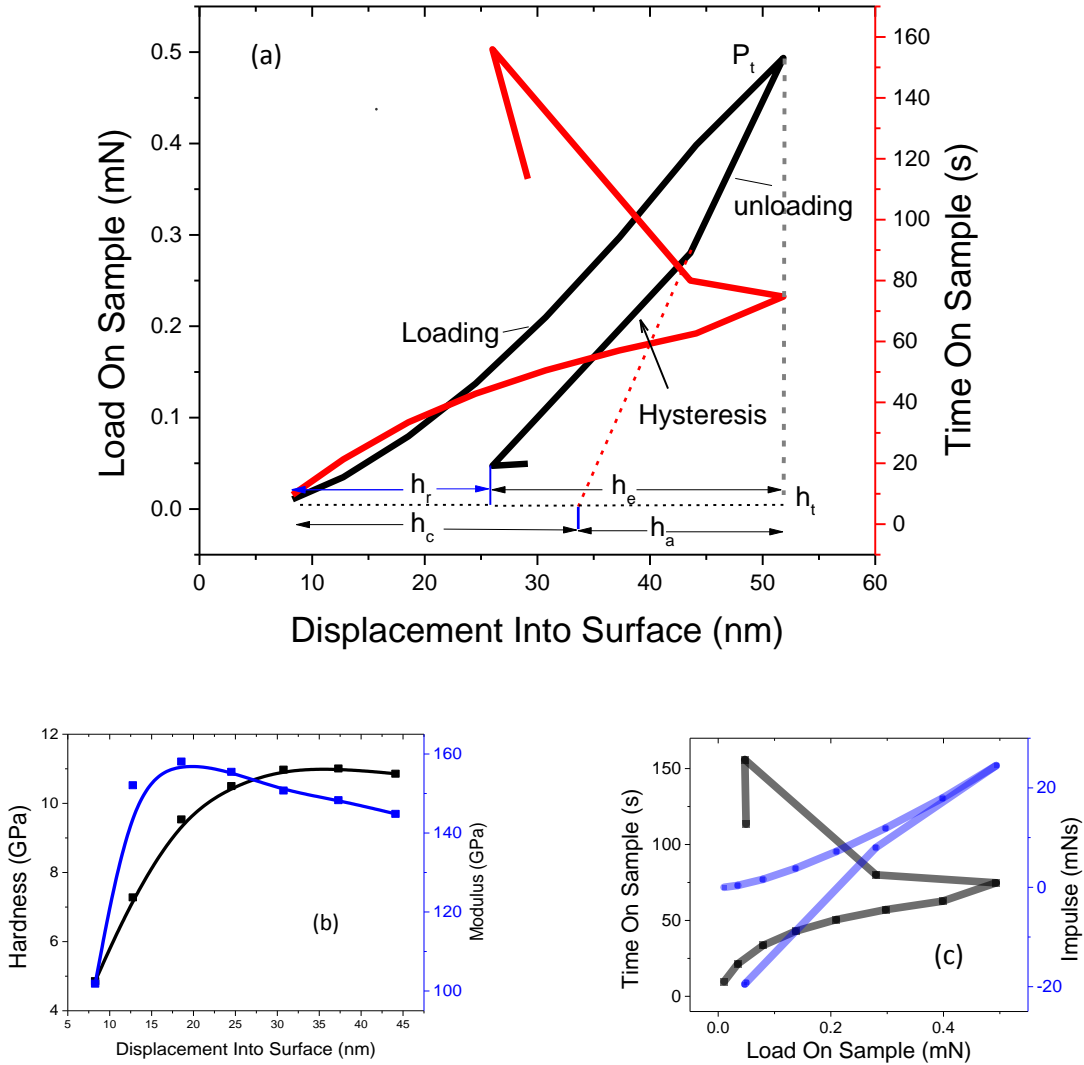

**Fig. 4**. Nanoindentation a) Load-depth (P-h) and Time on the sample b) Hardness and Elastic Modulus plots c) impulse w.r.t load and time on sample

Recent studies have also shown the influence of electric field on nanoindentation with an increase in maximum indentation depth and final penetration depth with an increasing electric field. The influence is due to competition between mechanical load and electric field in the domain switching process **[20]**. The fracture studies of materials used in Li batteries have been done through nanoindentation. Recent studies have also shown an effect of charge states on the yield and elastic modulus obtained through stress-strain plots based on nanoindentation using the power-law hardening model for Li$_x$Sn alloys **[21].**

Nanoindentation performed at low precession can be correlated to surface electrical properties. Current can pass through the contact region of the conducting indenter and surface. The resistivity offered to the current as a result varies with the forces applied causing plastic deformation. Plastic deformation has been a major contributor to resistivity which is very well established as the dislocation can as scattering centers for the electrons.

When two conductors come into contact, the resistance R of the contact region is made up of two parts: $R = R_c + R_f$, where $R_c$ is the constriction resistance that depends on the material's bulk properties and $R_f$ is the contact resistance brought on by the characteristics of surface layers. The contact region's constriction resistance is given by **eq 1 [22, 23].**

$$R_c = \frac{\rho_1+\rho_2}{2}\sqrt{\frac{\pi}{A}} \qquad (1)$$

where $\rho_1$ is the resistivity of the indenter material, $\rho_2$ is the resistivity of the sample material, and A is the contact area. If V is the voltage drop and I is the current in the contact region, then for an indentation depth $h$, we can have the following relation (**eq 2**) **[23].**

$$h\frac{V}{I} = \sqrt{\frac{\pi}{24.5}}\left(\frac{\rho_1+\rho_2}{2}\right) \qquad (2)$$

The resistivity of the diamond which is the indenter material is 0.1 Ω m. SiCN used for MEMS has been reported to have a room temperature resistivity of 5.5 Ω m which leads to the following expression h (V/I) = 1.89. V/I is nothing but the resistance which there is found to vary with indentation depth.

If the shape of the contact region is considered as a circle of radius a, then the constriction resistance is given as **eq 1**, where $\rho_1`$ is the resistivity due to tip and $\rho_2`$ is for the sample. For a circular region, the Rc is given as **eq 2**

$$R_c = \frac{\rho_1` + \rho_2`}{2a} \qquad (1)$$

$$R_c = \frac{\rho_1` + \rho_2`}{2}\sqrt{\frac{\pi H}{F}} \quad (2)$$

where H is the hardness and F is the force applied **[24]**. For a Berkovich indenter, the contact region is no longer circular and we take 24.5 h² as the contact area which modifies the resistance equation as **eq 3**.

$$R_c = \frac{\rho_1` + \rho_2`}{2h}\sqrt{\frac{\pi}{24.5}} = 0.18\frac{\rho_1` + \rho_2`}{h}. \quad (3)$$

The fractured region or the heavily plastically deformed regions should also influence the electrical properties as they indicate high strain fields surrounding the indentations which should interfere with the free electron flow motion. The force during nanoindentation and the current passing in the contact region vary with time **[24]**. A higher value of impulse is causing a lower current value hence a higher rate of force application causes an increased dislocation jamming causing a higher scattering of the conducting electrons. The impulse obtained from Load and Time-on-sample is given in **Fig 4(c).** The importance of impulse in nanoindentation has been reported **[25].**

Piezoresistive MEMS used in pacemakers is based on impulse received from the heart's vibration for energy production proving a longer lifetime of the device. A silicon-based MEMS cantilever is used which uses CMOS-compatible AlN as the piezoelectric layer and works on shock-induced vibration producing energy. SiCN having piezoelectric properties can replace AlN **[26]**. The mechanical response of N/MEMs under impulse loading is a major criterion for device fabrication **[27].** Cantilever-based MEMS resonator sensors suffer the problem of un attainment of homogenous mass sensitivity. The intensity of vibration being highest at the free end of the cantilever makes the mass sensitivity highest and reduced as one moves towards the fixed end. The resonant frequency of operation is inversely proportional to the mass., impulse plays a very significant role in these devices. Nanoindentation and stress analysis on cantilever nanobeams have been performed and reported **[28].**

High precession indentation at the nanoscale is a means of studying surface features, especially for materials used in MEMS applications. The capacitive controlled load-displacement features in a nanoindenter bring into consideration the conductivity phenomenon occurring at the contact point between the probe and the surface. The zero-point tip defects also influence the electrical properties as there occurs a difference in strain in the case of sharp factory default 3-sided pyramidal Berkovich tip contact and an effective hertzian contact due to tip blunting **[29]**. This stain is known to affect the resistivity as per Matthiessen's rule which talks about impurity scattering as one of the deciding factors and has correlations with the strain development at the tip-surface contact. The strain field gradients were determined to study the structure inhomogeneities and microstructure. The strain gradients have been found to be very crucial in the proper analysis of obtained values from nanoindentation testing **[29– 33]**. Elastic-plastic strain gradients are given as **eq 4 (a, b, c, d)** where *P'* represents the total strain gradient, *S'* represents the elastic total strain gradient, *E'* represents the elastic normal strain gradient, and *H'*, the plastic total strain gradient.

$$P' = \frac{h}{P}\frac{dP}{dh} \quad (4a)$$

$$S' = \frac{h}{S}\frac{dS}{dh} \quad (4b)$$

$$H' = \frac{h}{H}\frac{dH}{dh} \quad (4c)$$

$$E' = \frac{h}{E}\frac{dE}{dh} \quad (4d)$$

We found out the strain field gradients for a Nanoindentation performed on SiCN/Si substrates for the load(P), harmonic stiffness (S), hardness (H), and modulus (E) as given in **Fig 5 (a, b, c, d)** respectively. The films showed a a maximum hardness of 20 GPa and a modulus of 220 GPa under optimized conditions **[7, 8]**. The plots show linear and smooth characteristics of the strain gradients indicating the homogeneity of the microstructure. However, looking at the S' strain gradient at shallow depths some inhomogeneities can be observed which have been magnified and shown in **Fig 5e**. This deviation in elastic strain gradient occurs since at shallow depth a transition from elastic Hertzian spherical contact from the blunt tip to conical sides occurs. A clear demarcation was shown in the figure when the tip starts to penetrate the substrate. This increased sharpness leads to higher plastic deformation leading to the material getting strain hardened in a localized zone surrounding the indentation region. The dislocations piling up due to increased plastic deformation causes strain hardening.

An external bias is given to the indenter tip and sample surface to allow conduction to occur, as shown in **Fig 5f** with the current direction. The indentation and any related processes affect the constriction resistance ($R_C$) and the contact resistance $R_f$. **[34, 35]**. The red dotted lines represent current flows that constrict at the contact and are modified by indentation according to Matthiessen's rule, which states the contribution of deformation (plastic) in conduction. In addition to Rc, the resistance component affected by mechanical stimulus provided by the indenter is the resistance at the coating/substrate interface $R_{intf}$, which may not be active at shallow depths of indentation (less than 10% thickness). When employed as a conducting tip and as a MEMS nanoindenter with an integrated AFM cantilever gripper for nanomechanical characterizations such as transducers for biosensing, the tip voltage in Berkovich indenter will be substantially greater. **[36]**.

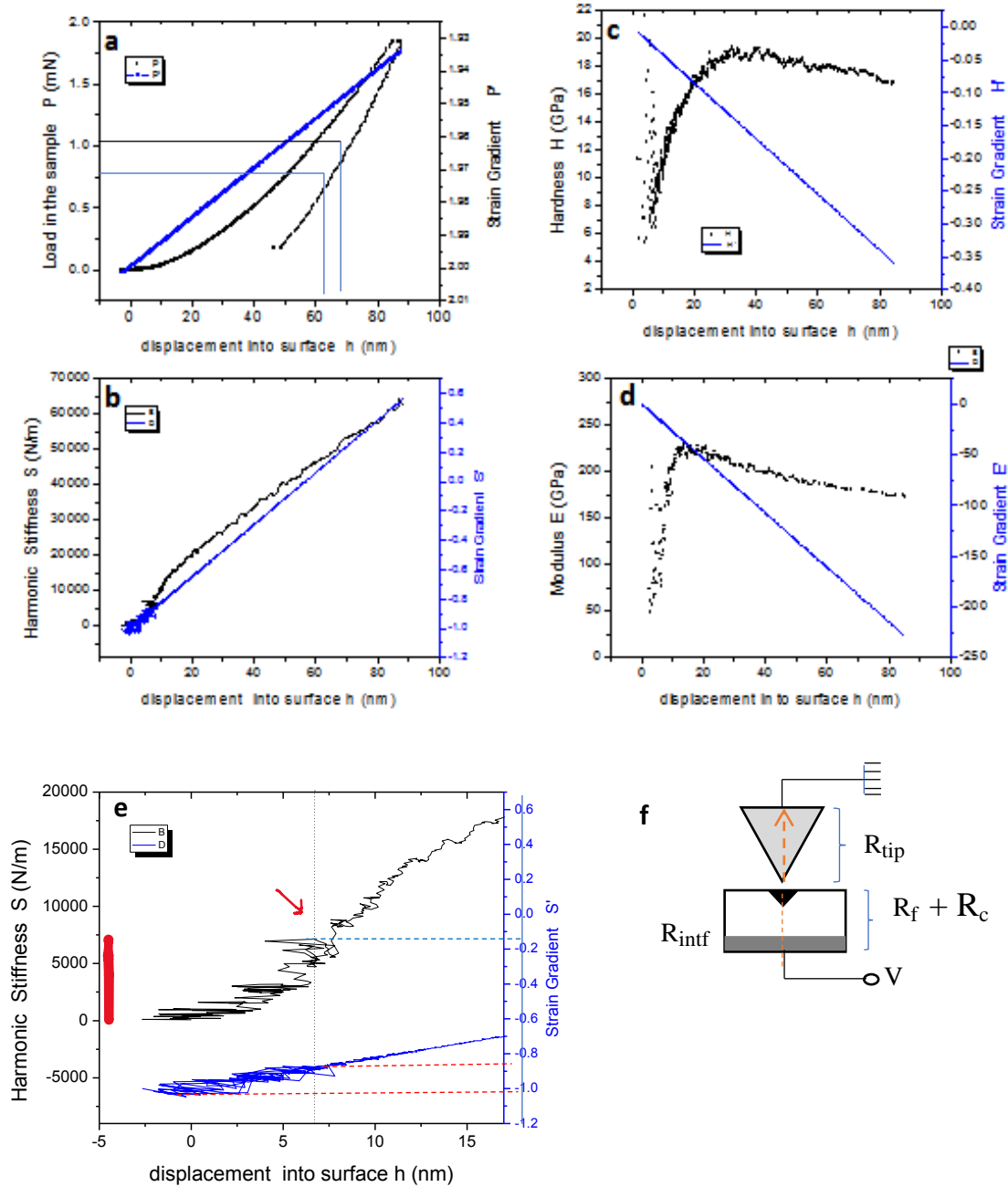

**Fig 5.** Strain Gradients corresponding to a) load-depth, b) Stiffness, c) hardness, and d) modulus. (e) Elastic Strain gradient at shallow depths (f) Electrical conduction between indenter tip and sample with different resistances **[34]**

From the substrate effect, it can be inferred that the coating thickness was about 200 nm as the substrate effect starts after 20 nm i.e 10% penetration (**Fig 5 c, d**). For a depth of about 6 nm, the resistance is 0.315 MΩ. According to the expression, there seems to be a linear decrease in R with h. However, there are other factors like plastic deformation, the effect of substrate, etc need to be considered before coming to any concrete statement. The strain gradient can be taken into consideration for this. For Harmonic stiffness change of 7500 N/m, the strain gradient shows a variation of 0.2 units for the initial contact. The sharpness of the tip has, therefore, a role to play in the tip-sample conduction mechanism. A Berkovich tip will have lesser strain gradient fluctuation at shallow depth and will be a path of enhanced conduction.

### 3.3 MEMS (µ)-cantilevers

MEMS nanoindenter with an integrated AFM cantilever gripper has been made for nanomechanical characterizations **[37]**. These micros (µ)-cantilevers act as transducers for biosensing. The cantilever response to biomolecule sensing depends on its mechanical properties determined by the spring constant and resonance frequency which are again functions of cantilever material and geometry. The change in stress occurring at the cantilever surface resulting in bending is expressed as **eq 5 [38]**.

$$\Delta\sigma = \frac{Et^2}{3(1-\nu)L^2}\Delta z \quad (5)$$

where E is Young's modulus, ν is the Poisson's ratio. L and t are the cantilever length and thickness respectively. Using the directional and total deformation values from **Fig 3(b, c)** and the dimension (**Fig 6a**) of the pedestal cantilever, the surface stress values for directional as well as total deformation were determined (**Table 2**). Both compressive and tensile stress may act on the structure creating changes in the curvature as shown in **Fig 6(b).** The silicon chip was 1mm thick with standard values of E =140 GPa and ν = 0.25. The deformations are given over 10 zones each having length 1.4 mm (= L) as the total dimension of one side of the structure is 14mm. If the system changes to SiCN/Si with E = 220 GPa, ν = 0.25 and we would like to find out the stress on SiCN/Si coating, the thickness of the system should be kept constant as the deformation is acting on the SiCN/Si system. So, the values just increase by a factor of 1.57 (220/140).

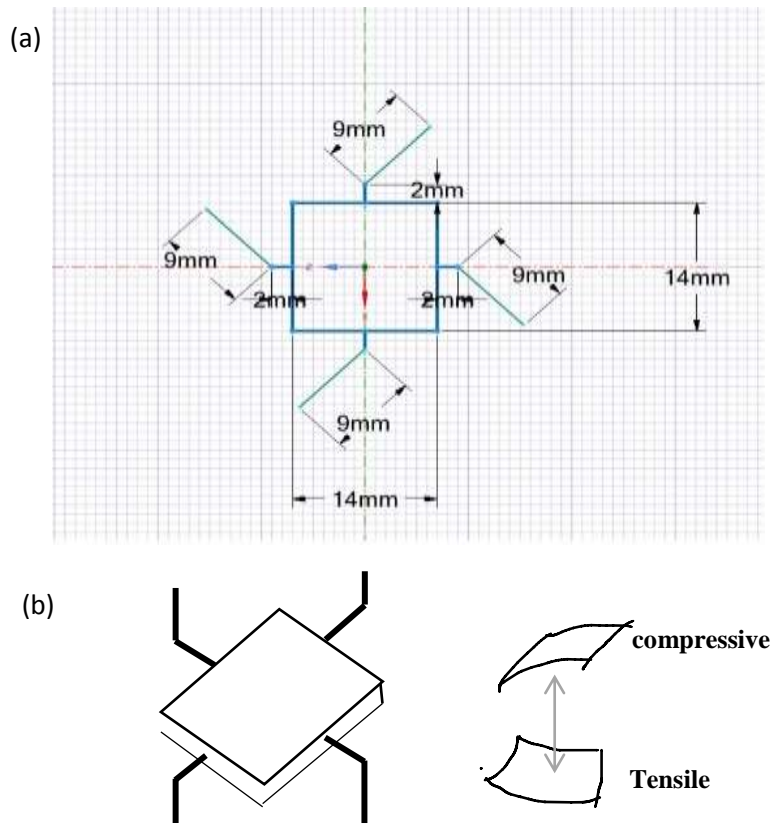

**Fig 6.** (a) The dimensions of the proposed pedestal cantilever structure and (b) vibrations associated with tensile and compressive stress

**Table 2.** The deformations in the pedestal cantilever structure and corresponding surface stress. (The positive values denote compressive stress and the negative values denote tensile stress)

| S. No | Δz (dir) (µm) | Δσ (dir) - Si (MPa µm) | Δz (T) (µm) | Δσ (T) (MPa µm) | Δσ (T) SiCN/Si (MPa µm) |
|---|---|---|---|---|---|
| 1 | 80.067 | 2.54 | 1.66 | 0.052 | 0.08164 |
| 2 | 60.675 | 1.92 | 1.47 | 0.046 | 0.07222 |
| 3 | 41.284 | 1.31 | 1.29 | 0.040 | 0.0628 |
| 4 | 21.892 | 0.69 | 1.10 | 0.034 | 0.05338 |
| 5 | 2.5005 | 0.08 | 0.92 | 0.028 | 0.04396 |
| 6 | -16.891 | -0.54 | 0.74 | 0.023 | 0.03611 |
| 7 | -36.283 | -1.15 | 0.55 | 0.017 | 0.02669 |
| 8 | -55.674 | -1.76 | 0.37 | 0.012 | 0.01884 |
| 9 | -75.006 | -2.38 | 0.18 | 0.006 | 0.00942 |
| 10 | -94.452 | -2.99 | - | - | - |

For coating substrate system, a modified form of the Stoney equation can be used as given in **eq 6 [39, 40]**

$$\sigma t = p \ln\left(\frac{KEh^2}{6p(1-v)} + 1\right) \text{ where } p = \frac{h^3}{D^2} E \left(\frac{13\cdot 5}{1-v^2} - 10.3\right) \quad (6)$$

*h* being the thickness of the total film/substarte system and t being the coating thickness. D is the diameter of the coating considering the shape to be circular in nature. For a film of 2 μm thickness and h being 1 mm. The pedestal cantilever being 14mm in one side can be considered as the value of D. The value of k, the spring constant is taken as the stiffness from the unloading portion of the load depth curve as shown in Fig 4a which comes around 0.35 MN/m. The value of p then comes out to be 150 GPa mm (or MPa m). Using this value of p the surface stress value comes out to be 711 MPa m.

**4. Summary**
The limitation of Si in high temperatures has caused alternative materials like SiC, $Si_3N_4$ etc to be used in small-scale mechano-electronic devices. MEMS-based Cantilever used for pressure sensors is one of the most important systems which are being fabricated using SiC-based materials. However, apart from the materials point of view, the geometry of the cantilever sensors required modifications to comply with the mass sensitivity effects. Finite Element Modeling of a novel design has therefore been provided which will help in the design of cantilever-based MEMS. The sharpness of the tip was found to influence the conduction occurring between the indenter and the sample material under consideration. The sharper the tip, the lesser will be any fluctuations occurring in strain gradients during the first contact. Prolonged use of a sharp 3-sided pyramidal Berkovich indenter causes tip blunting and therefore for shallow depth the contact is more Hertzian (Spherical) in nature which affects the conduction mechanism by causing an enhanced resistance. The surface stress was determined from Stoney's relation using the deformation obtained from the FEA.

ACKNOWLEDGEMENTS

The authors would like to thank the INUP program at CeNSE, IISc Bangalore, and Dr S. K. Mishra, CSIR - National Metallurgical Laboratory, for the instrumental facilities

DECLARATIONS

**Compliance with Ethical Standards**
The manuscript has not been submitted in parallel either in full or partially to any other journal.

**Conflict of interest**
There is no conflict of interest among the authors

**Research Data Policy and Data Availability Statements**
Data shall be provided on request

**Author Contribution**
All the authors have contributed equally to the paper

**FUNDING**
No funding was received for conducting the research